\documentclass[5p]{elsarticle}
\makeatletter
\def\ps@pprintTitle{%
 \let\@oddhead\@empty
 \let\@evenhead\@empty
 \def\@oddfoot{\centerline{\thepage}}%
 \let\@evenfoot\@oddfoot}
\makeatother

\usepackage[utf8]{inputenc}
\usepackage[english]{babel}
\usepackage[babel]{csquotes}
\usepackage{stmaryrd}
\usepackage{amssymb}
\usepackage{amsmath}
\usepackage{amsthm}
\usepackage{graphicx}

\providecommand{\R}{\mathbb{R}}

\begin{document}

\begin{frontmatter}
\title{The cosmological nature of the dark Universe}

\author{Christian Henke} 
\ead{henke@math.tu-clausthal.de}
\address{University of Technology at Clausthal, Department of Mathematics,\\ Erzstrasse 1, D-38678 Clausthal-Zellerfeld, Germany}

\date{\today}

\begin{abstract}
This paper deals with the cancellation mechanism, which identifies the energy density of space-time expansion in an empty universe with the zero-point energy density and avoids the scale discrepancy with the observed energy density (cosmological constant problem).
Using an intrinsic degree of freedom which describes the coupling of a variable cosmological term $\Lambda$ with non-relativistic matter and radiation, the following consequences are demonstrated by coupling only a small contribution of $\Lambda$ with non-relativistic matter. 
First, the standard model of cosmology with a positive cosmological constant can be generalised such that the missing mass problem of dark matter is solved by an overall negative variable cosmological term. 
Second, the model under consideration is compatible with constraints from the standard model of particle physics.
Third, an equation of state parameter of dark matter is derived which agrees with observations of rotational curves of galaxies.
Moreover, the creation and annihilation process of dark matter is presented.
\end{abstract}

\end{frontmatter}

\section{Introduction}
In the recent paper \cite{Henke_QuantumVacuumEnergy_2018}, the author has demonstrated that the variable cosmological term $\Lambda(a)=\Lambda_0+\Lambda_1 a^{-r},$ $r=4-\epsilon,$ $\epsilon=9.151 \cdot 10^{-122}$ solves the fine-tuning of the cosmological constant problem 
(see
\cite{Carroll_TheCosmologicalConstant_1992,Carroll_TheCosmologicalConstant_2001,Sahni.Starobinsky_TheCasefora_2000,Weinberg_TheCosmologicalConstant_1989,Dolgov_ProblemVacuumEnergy_1997})
and generates the missing mass of dark matter which constitutes $26\%$ of the matter-energy density. The remaining matter-energy constraints of the universe such as $69\%$ dark energy and $5\%$ ordinary matter as well as the initial singularity are also satisfied by the theory.   

Unfortunately, the age of the universe without contributions from radiation which is given by
\begin{equation*}
\frac{1}{H_0} \int_0^1 \frac{dx}{\sqrt{0.05/x + 0.69 x^2 +0.26/x^{r-2}}} =10.6 \cdot 10^9 \text{years},
\end{equation*}
disagrees with the accepted value of $13.80 \cdot 10^9$ years. 
Here, $H_0=67.74 \text{ km/(s Mpc)}$ denotes the present-day Hubble constant. 
The age of $13.80$ billion years can be obtained in the case $r \approx 3.$
In order to retain the accepted age of the universe, provided that the fine-tuning property of the cosmological constant problem and the generation of the missing mass of dark matter are preserved, an intrinsic degree of freedom for a parameterised coupling of the cosmological term with non-relativistic matter and radiation is used.
It turns out that in our universe, only a very small fraction of the cosmological term couples with non-relativistic matter. Interestingly, this is the same result as derived from the observational constraints for the baryon and antibaryon pair annihilation \cite{Freese.Adams.ea_CosmologyDecayingVacuumEnergy_1987}.
As another consequence, an equation of state parameter of dark matter is found which agrees with observations of rotational curves of galaxies
\cite{Barranco.Bernal.ea_DarkMatterEquationState_2015}.

The remainder of the paper is organised as follows. In section 2 we review the Friedmann equations where the cosmological term $\Lambda$ is a function of the scale factor $a$. Section 3 is devoted to the dark sector and states the basic ideas behind the derivation of the cancellation mechanism of the cosmological constant problem from \cite{Henke_QuantumVacuumEnergy_2018}.
Then, it is demonstrated that the attractive force of dark matter is a consequence of a positive energy density of the cosmological term. 
Finally, we show that our solution of the cosmological constant problem explains the already mentioned cosmological observations and how dark matter is created/destroyed by the interaction of non-relativistic matter/radiation.

\section{A time-dependent cosmological term}

Let us start by recalling the Robertson-Walker space-time line element 
(see \cite{Wald_GeneralRelativity_1984} for notational conventions)
\begin{equation*}
ds^2=-c^2 dt^2+a(t)^2 \left(\frac{1}{1-kr^2} dr^2+r^2 d \Omega^2 \right),
\end{equation*}
which reduces Einstein's field equations to the Friedmann equations
for the scaling factor $a(t)$
\begin{align}
\frac{\dot{a}^2}{a^2} -  \frac{1}{3} \Lambda +\frac{k}{a^2}&= \frac{\kappa c^2}{3}  \rho,
\label{eq:friedmann0}\\
3\frac{ \ddot{a}}{a} - \Lambda  &=-\frac{\kappa}{2} \left( \rho c^2 + 3p \right).
\label{eq:friedmann1}
\end{align}
Here, 
$d \Omega^2=d\theta^2 + \sin^2 \theta d \phi^2$ and 
$k$ denotes the curvature parameter of unit $\text{length}^{-2}.$ 
Moreover, the variable cosmological term is defined by 
\begin{equation}
\Lambda(a)=\Lambda_0 + \Lambda_1 a^{-r},\quad \Lambda_0,r>0.
\label{eq:lambda_model}
\end{equation}
It is convenient to include the $\Lambda$-term in the energy-mo\-men\-tum tensor of the right-hand side. Therefore, we define an effective density and pressure field which includes non-relativistic matter and radiation
\begin{equation*}
\rho_{\text{eff}}=\rho_m + \rho_r + \rho_\Lambda, \quad p_{\text{eff}}=p_m + p_r +p_\Lambda,
\label{eq:eff_rho_p}
\end{equation*}
where $\rho_\Lambda=\Lambda(a)/\kappa c^2$ and $p_\Lambda=-\Lambda(a)/\kappa$ denote the background fields. 

The zero covariant divergence of the energy-momentum tensor gives
\begin{equation}
\frac{d}{da} \left(\rho_{\text{eff}} a^3 \right) +3 \frac{p_{\text{eff}}}{c^2} a^2=0, \quad p'_\text{eff}=0,
\label{eq:divT_rho_p}
\end{equation}
and leads with $p_m=0$ and $p_r=c^2 \rho_r/3$ to
\begin{equation}
\frac{1}{a^3}\frac{d}{da} \left(\rho_m a^3 \right) 
+\frac{1}{a^4}\frac{d}{da} \left(\rho_r a^4 \right) 
=- \frac{1}{\kappa c^2} \frac{d}{da} \Lambda(a),
\label{eq:divT_rho_p2}
\end{equation}
which is satisfied by the parametrised coupling
\begin{align}
\frac{1}{a^3}\frac{d}{da} \left(\rho_m a^3 \right) 
&=- \alpha \frac{1}{\kappa c^2} \frac{d}{da} \Lambda(a),\\
\frac{1}{a^4}\frac{d}{da} \left(\rho_r a^4 \right) 
&=- (1-\alpha) \frac{1}{\kappa c^2} \frac{d}{da} \Lambda(a),\quad \alpha \in \R.
\end{align}
Integrating both equations lead to
\begin{align}
\rho_m&= \rho_{m,0} a^{-3}+ \frac{\alpha r \Lambda_1}{\kappa c^2}\frac{1}{3-r} a^{-r},\quad r \neq 3,\label{eq:rho_m}\\
\rho_r&= \rho_{r,0} a^{-4}+ \frac{(1-\alpha) r \Lambda_1}{\kappa c^2}\frac{1}{4-r} a^{-r},\quad r \neq 4.\label{eq:rho_r}
\end{align}
Hence, depending on the exponent $r,$ the presence of the variable cosmological term can completely change the dynamical behavior of $\rho_m$ and $\rho_r.$

\section{The dark sector}
In this section we review the basic ideas of the cancellation mechanism between the quantum zero-point energy and the energy density of the cosmological constant in an empty universe (cf. \cite{Henke_QuantumVacuumEnergy_2018}). First, using a metric which is independent of the scale factor, the second Friedmann equation can be identified with an Euler-Lagrange equation. Therefore, the related Lagrangian and the first Friedmann equation leads to the cancellation mechanism 
between the energy density of space-time expansion and the energy density of the cosmological term. 
Identifying the energy density of space-time expansion with the quantum zero-point energy, 
we get 
\begin{equation}
\rho_{\Lambda}= \rho_{\text{zpe}}+\frac{4 \int^a \Lambda(\alpha) \alpha^3 \, d\alpha}{\kappa c^2 a^4},
\label{eq:cancellation_mech}
\end{equation}
where
\begin{equation}
\rho_{\text{zpe}}=\frac{\Lambda_1}{\kappa c^2} \frac{r}{r-4} a^{-r}, \quad
\Lambda_1=\frac{r-4}{2 \pi r l_p^2},\quad r\neq 4,
\label{eq:lam1}
\end{equation}
and $l_p$ denotes the Planck length.
As the consequence, there is no scale discrepancy between the total energy densities and the fine-tuning problem of the cosmological constant problem is solved.

Now, we investigate the parameter range for $r$ such that an initial singularity is guaranteed. Using the settings 
\begin{align*}
\rho_{\Lambda_1}&= \frac{\Lambda_1}{\kappa c^2} a^{-r},\\
\rho_{m\Lambda}&=\frac{\alpha \Lambda_1}{\kappa c^2}\frac{r}{3-r} a^{-r},\quad r \neq 3,\\
\rho_{r\Lambda}&=\frac{(1-\alpha) \Lambda_1}{\kappa c^2}\frac{r}{4-r} a^{-r},\quad r \neq 4,
\end{align*}
we have to discuss the inequality
\begin{equation*}
\rho_{\Lambda_1}+\rho_{m\Lambda}+\rho_{r\Lambda}
=\frac{\Lambda_1}{\kappa c^2}\left( 1+\frac{\alpha r}{3-r}+\frac{(1-\alpha)r}{4-r} \right) a^{-r}\geq 0. 
\end{equation*}
By considering the equation 
\begin{equation*}
\frac{r-4}{r} \left( 1+\frac{\alpha r}{3-r}+\frac{(1-\alpha)r}{4-r}\right)
=\frac{\left(4-\alpha\right)\left(r-\frac{12}{4-\alpha}\right)}{r(3-r)},
\end{equation*}
we find that the inequality is satisfied if
\begin{equation}
3 < r < \frac{12}{4-\alpha}.
\label{eq:param_range_r}
\end{equation}
In order to analyse the gravitational nature of the $\Lambda$-de\-pen\-ding components of $\rho_{\text{eff}},$ the acceleration behaviour is discussed by the term
\begin{equation}
\begin{split}
&-\frac{\kappa c^2}{2}\left( \rho + 3\frac{p}{c^2} \right)+ \Lambda
\\
&=-\frac{\kappa c^2}{2}\left( \rho_{m,0} a^{-3}  + 2 \rho_{r,0} a^{-4} \right)+ \Lambda_0\\
&- \Lambda_1 \left( -1 + \frac{\alpha r}{2(3-r)} +\frac{(1-\alpha)r}{4-r} \right) a^{-r}, \quad 3 \neq r \neq 4,
\end{split}
\label{eq:accel_beh}
\end{equation}
from equation $(\ref{eq:friedmann1}).$ 
As usual, the $\Lambda_0$-depending term acts repulsive and is labeled with dark energy.
Since 
\begin{equation*}
\begin{split}
&\phantom{=,}\frac{r-4}{r} \left( -1+\frac{\alpha r}{2(3-r)}+\frac{(1-\alpha)r}{4-r}\right)\\
&=\frac{\left(4-\alpha\right)\left(r-\frac{12}{4-\alpha}\right)\left( r-2 \right)}{2r(3-r)},
\end{split}
\end{equation*}
it follows from the considered parameter range $(\ref{eq:param_range_r})$ that the $\Lambda_1$-depending term has an attractive effect and can be identified as dark matter. Notice that the attractive nature of dark matter is a direct consequence of a positive $\Lambda_1$-depending density.

From $(\ref{eq:rho_m})$ and $(\ref{eq:rho_r})$ we can conclude that dark matter doesn't interact with ordinary matter and radiation ($\rho_{m,0}$ and $\rho_{r,0}$ are independent of $\Lambda_1$). The interaction takes place between the three components of dark matter (self-interacting dark matter) which can be classified by their equation of state parameter $w=-1, w=0$ and $w=1/3.$ 

\section{Cosmological constraints}
It remains to consider some observational constraints for the present-day composition of the universe. Using the usual settings $\rho_{\text{crit}}=3 H_0^2/\kappa c^4,$
\begin{equation}
\begin{aligned}
\Omega_k&=&-\frac{k c^2}{a_0^2 H_0^2}&,& 
\Omega_m&=& \frac{\rho_{m,0}}{\rho_{\text{crit}}} a_0^{-3},\\
\Omega_r&=& \frac{\rho_{r,0}}{\rho_{\text{crit}}} a_0^{-4}&,&
\Omega_{\Lambda_0}&=&\frac{\Lambda_0 c^2}{3 H_0^2},
\end{aligned}
\end{equation}
where $a_0=a(t_0)$ denotes the present-day scale factor, the dark matter contribution is taken into account by
\begin{equation}
\Omega_{dm}=\frac{\rho_{\Lambda_1}+\rho_{m\Lambda}+\rho_{r\Lambda}}{\rho_{\text{crit}}}.
\end{equation}
Multiply $(\ref{eq:friedmann0})$ by $c^2 a^2/H_0^2 a_0^2,$ we get 
\begin{equation}
\left( \frac{dx}{d\tau} \right)^2=\Omega_k +\frac{\Omega_m}{x} + \frac{\Omega_r}{x^2} +\Omega_{\Lambda_0} x^2 +\frac{\Omega_{\text{dm}}}{x^{r-2}},
\label{}
\end{equation}
where $x=a/a_0$ and $\tau=H_0 t.$ This leads to the present-day constraint
\begin{equation}
1=\Omega_k + \Omega_m + \Omega_r + \Omega_{\Lambda_0}+ \Omega_{\text{dm}}.
\end{equation}
To relate the last equation with observations (cf. \cite{Ade_Planck_2015}), we consider 
\begin{equation*}
(\Omega_k,\Omega_m,\Omega_r,\Omega_{\Lambda_0},\Omega_{dm})=(0,0.05,5\cdot 10^{-5},0.69,0.26)
\end{equation*}
and $H_0=67.74 \frac{km}{s\, Mpc}.$ 
The setting $\Omega_{\Lambda_0}=0.69$ leads to a cancellation mechanism $(\ref{eq:cancellation_mech})$ which automatically cancel 121 decimal places without fine-tuning.
Moreover, using
$\Omega_{\text{dm}}=0.26$ and
\begin{equation*}
\epsilon=\frac{2 \pi l_p^2 H_0^2 \Omega_{\text{dm}}}{c^2}=2.29 \cdot 10^{-122},
\end{equation*}
it follows
\begin{equation}
3 \epsilon r (3-r)=r(4-\alpha)-12,
\label{eq:sing_perturbed}
\end{equation}
which is a singular perturbed equation and can be solved by an uniform asymptotic expansion
\begin{equation}
r(\epsilon)=r_0+r_1 \epsilon + \mathcal{O}(\epsilon^2),
\label{eq:r_asymp}
\end{equation}
where $r_0$ and $r_1$ are independent of $\epsilon.$ By substituting the decomposition $(\ref{eq:r_asymp})$ in $(\ref{eq:sing_perturbed})$, a hierarchy of equations for the terms multiplied by the same power of $\epsilon$ follows and yields the solution to the unknowns $r_0$ and $r_1.$
Namely, we obtain
\begin{equation}
r=3 +\frac{3 \alpha}{4-\alpha} -\frac{108 \alpha \epsilon}{(4-\alpha)^3} + \mathcal{O}(\epsilon^2).
\label{eq:r3}
\end{equation}
As already mentioned in the introduction a small coupling between the non-relativistic matter and the dynamical part of the cosmological term is necessary for an universe which is $13.80$ billion years old.

\begin{figure}[!htbp]
\begin{center}
\includegraphics[width=1.0\columnwidth]{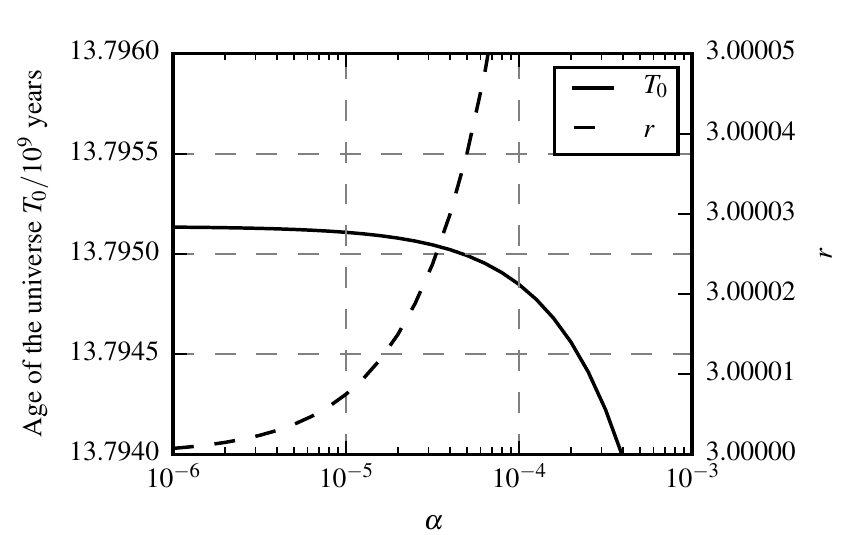}
\caption{The age of the universe $T_0$ and the dark matter exponent $r$ with respect to the coupling parameter $\alpha$.}
\label{fig:age_universe_alpha}
\end{center}
\end{figure}

Namely, using $(\ref{eq:r3})$ with $0<\alpha<10^{-5}$ we get by numerical integration 
\begin{equation*}
\frac{1}{H_0} \int_0^1 \frac{dx}{\sqrt{\frac{\Omega_m}{x} + \frac{\Omega_r}{x^2}+\Omega_{\Lambda_0}x^2 +\frac{\Omega_{\text{dm}}}{x^{r-2}}}} =13.80 \cdot 10^9 \text{years}.
\end{equation*}
Therefore, the remaining parameter of the cosmological term is $\Lambda_1=-2.03 \cdot 10^{68}.$ Hence, the overall cosmological term $\Lambda(a)$ is negative! (cf. (\ref{eq:lam1}) and Figure \ref{fig:age_universe_alpha}).
Now, some consequences of the value $0<\alpha<10^{-5}$ are investigated.
The low production $(\rho_{m\Lambda}>0)$ with respect to $\alpha$ of non-relativistic (dark) matter particles by the coupling process between non-relativistic matter and the cosmological term is in agreement to the results of \cite{Freese.Adams.ea_CosmologyDecayingVacuumEnergy_1987}:
the limitation of baryon/an\-ti\-baryon pair annihilations, the isotropy of the microwave background and equation $(\ref{eq:divT_rho_p2})$
lead to a scale factor which is nearly indistinguishable to the scale factor in the matter dominated epoch. Hence, the term $\frac{d}{da} (\rho_m a^3)$ is only a small perturbation of equation $(\ref{eq:divT_rho_p2}),$ which is equivalent to the case $0 < \alpha \ll 1.$

Interestingly, the amount of annihilation of dark matter corresponds to the amount of the zero-point energy (cf. \cite{Henke_QuantumVacuumEnergy_2018})
\begin{equation*}
\rho_{r\Lambda}=-(1-\alpha) \rho_{\text{zpe}} \approx - \rho_{\text{zpe}}.
\end{equation*}

\begin{figure}[!htbp]
\begin{center}
\includegraphics[width=1.0\columnwidth]{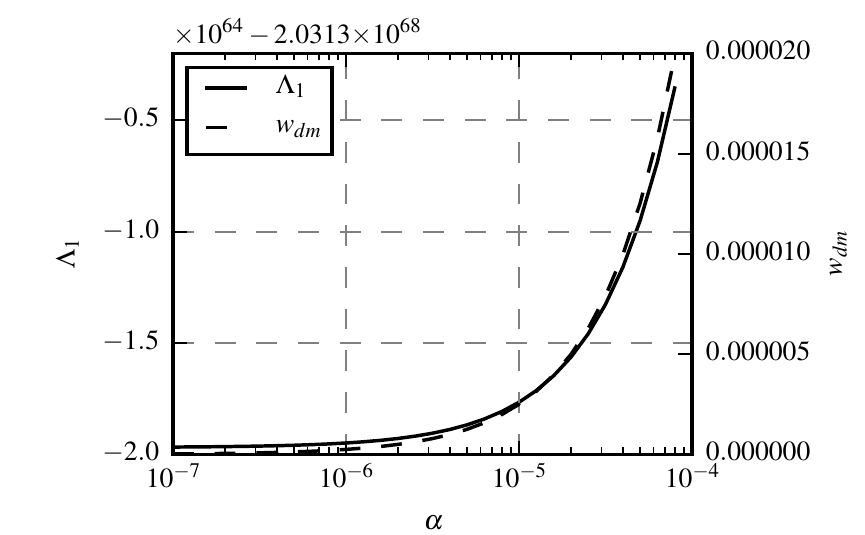}
\caption{The cosmological parameter  $\Lambda_1$ and the equation of state parameter $w_\text{dm}$ with respect to the coupling parameter $\alpha$.}
\label{fig:lambda_wdm_alpha}
\end{center}
\end{figure}

Finally, the total equation of state parameter of dark matter is given by
\begin{equation*}
\begin{split}
w_{\text{dm}}&=\frac{\frac{1}{3} \rho_{r\Lambda}-\rho_{\Lambda_1}}{c^2\left( \rho_{\Lambda_1}+\rho_{m\Lambda}+ \rho_{r\Lambda} \right)}
=\frac{r}{3}-1 \\
&=\frac{\alpha}{4-\alpha}+ \mathcal{O}(\epsilon)
\le 2.5 \cdot 10^{-6},
\end{split}
\end{equation*}
which agrees with the equation of state parameter $w_{\text{dm}}\in (1.3 \cdot 10^{-8}, 1.5 \cdot 10^{-7})$ 
from observed rotational curves of galaxies (see \cite{Barranco.Bernal.ea_DarkMatterEquationState_2015})(Figure \ref{fig:lambda_wdm_alpha}).
In contrast to the standard $\Lambda\text{CDM}$ model with $6$ parameters, the equation of state parameter $w_\text{dm}$ extends this model to the  $\Lambda\text{wDM}$ model from \cite{Thomas.Kopp.ea_ConstrainingPropertiesDarkMatter_2016}.
The above constraint for $w_\text{dm}$ increases the Hubble constant by a maximum of 0.007\% and decreases $\Omega_\text{dm}$ by a maximum of 0.02\% (cf. \cite[Figure 7]{Thomas.Kopp.ea_ConstrainingPropertiesDarkMatter_2016}) which is within the specified accuracies of the parameters.
Moreover, the derived equation of state parameter $w_\text{dm}$ meets the constraint $-0.000896 < w_\text{dm}<0.00238$ from cosmic microwave background observations \cite{Thomas.Kopp.ea_ConstrainingPropertiesDarkMatter_2016}.

The structure formation of the universe for the $\Lambda\text{wDM}$ model was investigated in \cite{Hu_StructureFormationGeneralizedDarkMatter_1998}. The main result is that the clustering scale of the large-scale structure is independent of the equation of state parameter. 

\section{Concluding remarks}
In this paper, the variable cosmological term $\Lambda(a)=\Lambda_0 + \Lambda_1 a^{-r}, r >0$ has been applied and it has been confirmed that the total energy density of an empty Friedmann universe is related to the cosmological term such that the fine-tuning problem was avoided by setting $\Lambda_1=r-4/2 \pi r l_p^2, r \neq 4.$ As a consequence, the dynamical part of the cosmological term generates the attractive force of dark matter.

Moreover, it has been demonstrated that the accepted age of our universe requires that only a small fraction of the cosmological term couples with non-relativistic matter. More precisely, the cosmological term creates/destroys dark matter with the interaction of non-relativistic matter/ra\-di\-ation.
Similar to black holes, the decrease of the density by radiation is caused by the vacuum energy density. This could be a further argument that dark matter consists of black holes and that $\rho_{\text{dm}}$ represents the spatially averaged densities of black holes.
On the other hand, the results of this paper cannot exclude that dark matter is made of particles. 
The range $3<r<3.0000075$ is compatible with constraints from baryon/antibaryon pair annihilation \cite{Freese.Adams.ea_CosmologyDecayingVacuumEnergy_1987}.

Furthermore, the parameter range of $r$ produces a non-zero equation of state parameter of dark matter $w_{\text{dm}} \le 2.5 \cdot 10^{-6}$ which agrees with observational data of rotational curves of galaxies \cite{Barranco.Bernal.ea_DarkMatterEquationState_2015},
generates the missing mass of dark matter,
describes the present-day composition of the universe
and realises a negative cosmological term. 
Living in an universe with a negative cosmological term will completely change our understanding of cosmology.
This could have important consequences for holographic correspondence-theories which are mainly formulated on space-times with a negative cosmological constant (cf. \cite{Aharony.Gubser.ea_LargeNField_2000}).



\end{document}